\documentclass[
    ,final            
  ]
  {aipproc}

\layoutstyle{6x9}

\renewcommand\XFMtitleblock{%
  \XFMtitle
  \let\XFMoldpar\par
  \def\par{\XFMoldpar\def\par{\space 
      for the VERITAS Collaboration\footnote{\protect\url{http://veritas.sao.arizona.edu/}}\XFMoldpar}}%
   \XFMauthors
   \let\par\XFMoldpar
   \XFMaddresses
   \XFMabstract
   \vspace{5pt}%
   \XFMkeywords
   \XFMclassification
}

\begin{document}

\title{Observations of very high energy emission from B2~1215+30 with VERITAS}

\classification{98.54.Cm; 95.85.Pw; 98.70.Rz; 98.54.-h}
\keywords      {AGN, gamma rays, non-thermal, VERITAS}

\author{Heike Prokoph}{
  address={DESY, Platanenallee 6, 15738 Zeuthen, Germany},
}

\begin{abstract}
We present results of 45 hours of data taken from January to June
2011 with VERITAS above 200~GeV in the direction of B2~1215+30. The blazar was
clearly detected with a constant flux of 
$F(E>200 \mathrm{GeV}) = (8.0 \pm 0.9) \times 10^{-12} \, \mathrm{cm}^{-2}
\mathrm{s}^{-1}$. 
The energy spectrum can be fitted using a power-law with index 
$-3.6 \pm 0.5$. 
Simultaneous multi-wavelength data are also presented, and are used to 
model the spectral energy distribution with a one-zone leptonic
model. 
\end{abstract}

\maketitle


\section{Introduction}
Blazars (BL Lac objects and flat spectrum radio quasars) 
are the most extreme class of active galactic nuclei. Their jet is
aligned closely with the line-of-sight and they show variability at
all wavelengths and time scales. 
One of the first BL Lac type objects to be identified was B2~1215+30 
\cite{Browne1971} (also known as 1ES~1215+303 or ON~325). 
Its redshift is uncertain with two different
values found in literature: z~=~0.130 \cite{Akiyama2003} and z~=~0.237
\cite{White2000}. 

At very high energies (VHE; E$>$100 GeV) B2~1215+30 was first detected 
by MAGIC in 2011 \cite{Mariotti2011}. 
VERITAS also observed the source, which is 
in the same field of view as two other VHE blazars: 1ES~1218+304 and
W~Comae. The observations of B2~1215+30 with VERITAS in 2011 are
reported on here.

\section{VERITAS Observations and Results}

VERITAS is an array of four imaging atmospheric Cherenkov telescopes 
located in southern Arizona, sensitive to gamma-ray
energies from 100 GeV to about 30 TeV, with a field of view of
$3.5^{\circ}$. 
For more details on the
VERITAS instrument see \cite{Galante2012}. 

VERITAS observed the sky towards B2~1215+30 for more than 45 hours
between January and June 2011, including dedicated pointings and
observations of 1ES~1218+304, which is only $0.76^{\circ}$ away. 
After quality selection and acceptance correction, a total of 38 hours
of data remained. 
The data were analyzed with 
pre-defined cuts, optimized for a 5\% Crab Nebula-like source. 
The signal was extracted using a ring
background model with an ON region of $\theta^2$ radius of $0.008
\mathrm{deg}^2$ centered on the position of B2~1215+30. 

In this ON region, 193 excess events have been detected with a
significance of $10.4 \sigma$ (using Eq. 17 in \cite{LiMa1983}). 
The flux above 200~GeV is found to be 
$(8.0 \pm 0.9) \times 10^{-12}\, \mathrm{cm}^{-2} \mathrm{s}^{-1}$, 
corresponding to about 3\% of the Crab Nebula flux above the same
energy threshold. 
The derived differential photon spectrum can be fitted by a power law: 
dN/dE = $F_0 (E/E_0)^{-\alpha}$, 
with $F_0 = (2.3 \pm 0.5) \times 10^{-11}$ cm$^{-2}$ s$^{-1}$, 
$E_0$ = 300~GeV, and $\alpha = 3.6 \pm 0.5$. 
This is compatible with the 2011 MAGIC results reported in
\cite{Aleksic2012}.

\section{Multi-wavelength observations}

Simultaneous multi-wavelength data have been taken by 
{\it Fermi}-LAT in the high energy regime, {\it Swift}-XRT in X-rays,
{\it Swift}-UVOT in the UV/optical, and with Super-LOTIS and MDM in the
optical. 

\begin{figure}[tb]
  \centering
  \includegraphics[width=1.0\textwidth]{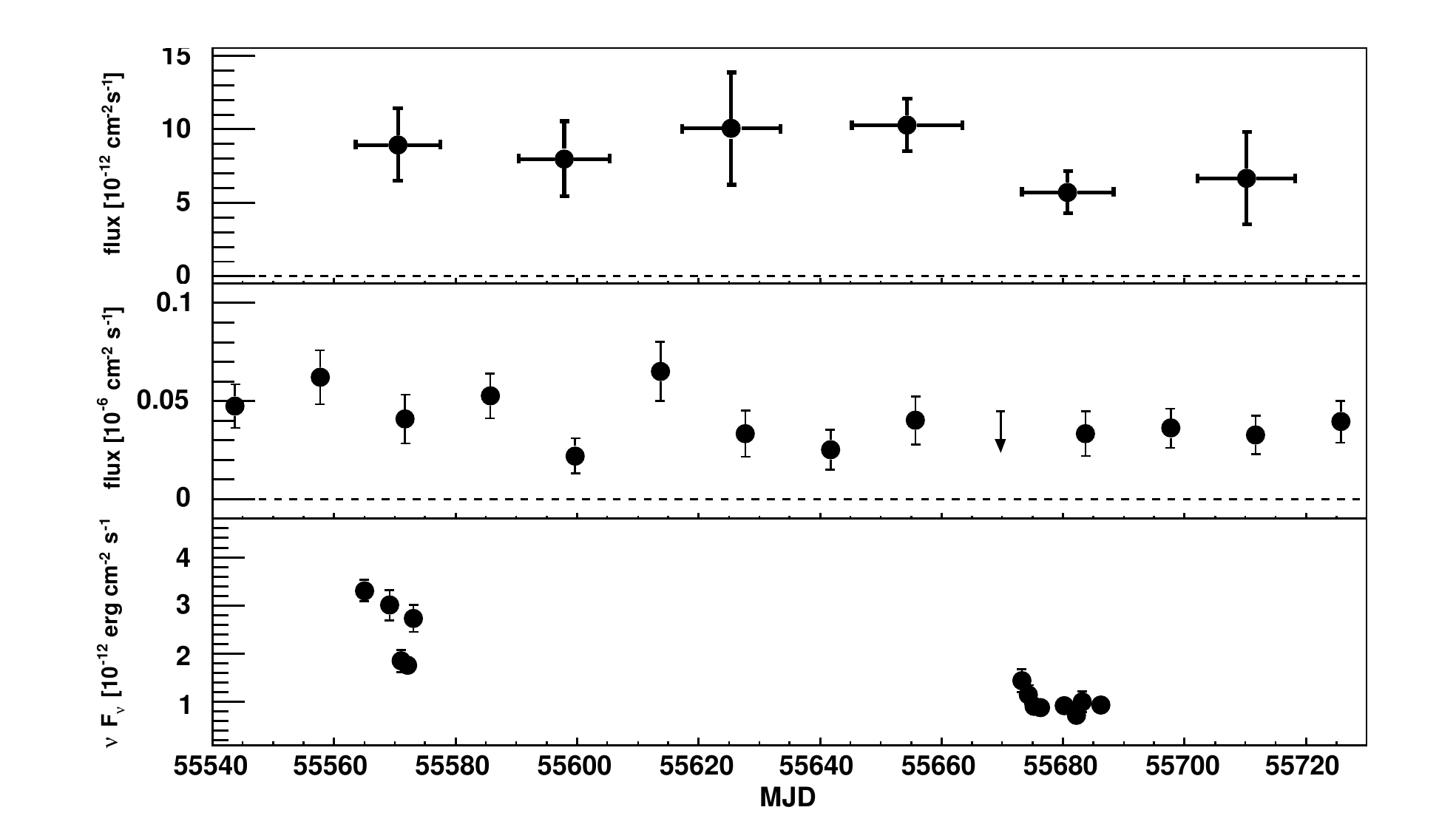}
  \caption{Light curves of B2~1215+30 for the first
    half of 2011: 
    (top) monthly binned VERITAS light curve (E$>$200~GeV),
    (middle) bi-weekly {\it Fermi}-LAT light curve (E$>$200~MeV), 
    (bottom) {\it Swift}-XRT light curve (2-10~keV). 
  }
  \label{fig:LC}
\end{figure}

The {\it Fermi}-LAT analysis was performed using the standard LAT
ScienceTools (version v9r23p1) and P7SOURCE\_V6 instrument response
functions together with the recommended cuts and 
background files. A binned maximum-likelihood method was applied to
the data in order to extract the light curve (shown in Fig. \ref{fig:LC}) and
the spectrum for the first half of 2011. 

{\it Swift}-XRT observed B2~1215+30 during January and 
in March/April 2011. The analysis was performed with HEASoft, XSPEC
v12.6.0 and the recommended response files. The spectrum was fitted
with an absorbed power law while the Galactic column density was fixed
to $N_H = 1.74 \times 10^{20}\, \mathrm{cm}^{-2}$ \cite{Kalberla2005}. 

In addition to the {\it Swift}-UVOT data, MDM took data
in May 2011 in the R, I and V band, 
while Super-LOTIS monitored the variability of B2~1215+30 in the R-band. 
The data reduction of the optical and UV data sets followed standard
methods. 

In Fig. \ref{fig:LC} the light curves for the first half of 2011 are
shown. 
In the VHE regime, no variability could be detected in the monthly
binned VERITAS light curve for energies above 200~GeV 
($\chi^2/ndf = 4.7/5$). 
Even though variability has been measured 
by {\it Fermi}-LAT in the past \cite{2FGL}, the bi-weekly binned light
curve above 200~MeV contemporaneous to the VERITAS observation is
compatible with being constant. 
In X-rays, 
the blazar is found to be brighter and harder in January than in
March/April 2011. Based on those different X-ray states, two SEDs have
been extracted which are shown in Fig. \ref{fig:SED}. 

For the January SED the highest X-ray flux (on MJD 55565) was used,
together with simultaneous UVOT data. 
To represent the low X-ray state of the source in Mar/Apr, a combined fit of
the XRT data was performed (using all available data from MJD
$55673-55686$). This was accomplished by simultaneous UVOT data and 
quasi-simultaneous MDM data from May 2011. 
Since no variability was detected in the high-energy regime in 2011, 
contemporaneous data from VERITAS and {\it Fermi}-LAT are used (MJD
$55560-55720$). 
Radio data were retrieved using the NED website \cite{NED}.

\section{Modeling of the Spectral Energy Distribution}

The extracted quasi-simultaneous SEDs are modeled with a synchrotron
self-Compton (SSC) model \cite{BoettcherChiang2002}, which takes the
intergalactic absorption into account using the model of Finke et
al. \cite{Finke2010}. 
Within the model the low-energy emission (from radio to X-rays) is
interpreted as 
synchrotron emission from relativistic electrons in a spherical
emission region, moving relativistically along the jet, which is
closely aligned with the line-of-sight. The high-energy gamma-ray
emission is produced via Compton up-scattering off the same electron
population which produced the synchrotron peak. 
The non-thermal electron distribution in the emission region is
determined self-consistently as an equilibrium between injection of a
power-law distribution with index $\alpha_e$, radiative cooling, and
escape. 

\begin{figure}
  \centering
  \includegraphics[width=0.9\textwidth]{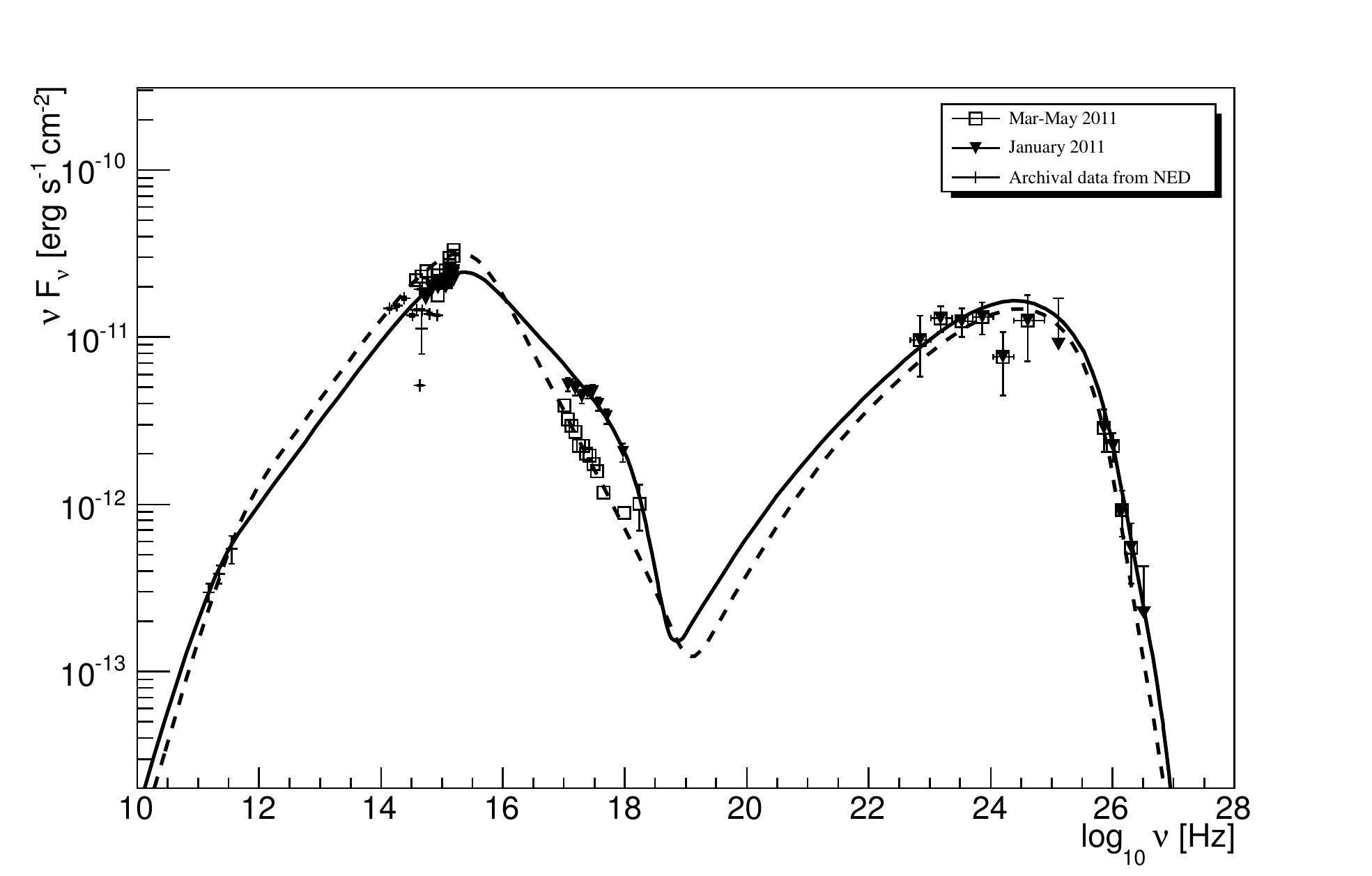}
  \caption{Spectral energy distribution of B2~1215+30. The solid line
    is the SSC model for the Jan 2011 data set, the dashed line
    represents the model for Apr 2011 incl. MDM data from
    May. }
  \label{fig:SED}
\end{figure}

The SEDs are well described by the SSC model (see Fig. \ref{fig:SED}). 
For a redshift z~=~0.130 the bulk Lorentz factor is $\Gamma$~=~30
while the emission region radius is $\sim 10^{17} \mathrm{cm}$. 
The two different flux levels in X-ray are accounted for by changing
the electron injection spectral index and the magnetic field strenght. 
The magnetic field stranght is $0.02$~G for January and $0.01$~G for
April, respectively. 
The electron injection spectral index is $\alpha_e = 2.8$ (Jan) and
$\alpha_e = 3.4$ (Apr) and is tightly constrained by the X-ray
spectrum. 

If relativistic shock acceleration takes
place within the jet, the change of the
electron distribution may be explained by a
change in the shock field obliquity \cite{SummerlinBaring2012}.
Within the modeling, the different X-ray states lead also to flux
variations in the high-energy peak. Nevertheless, those
could not been detected by {\it Fermi}-LAT or VERITAS given their
sensitivity within the observation period reported here. 

For z~=~0.237 the SSC modeling is more challenging. Indeed, a higher Lorentz
factor ($\Gamma$~=~50) is required and the predicted VHE flux is below
the VERITAS measurement. However, given the statistical and systematic
errors on the energy spectrum of the high-energy peak, this redshift
cannot be excluded within the framework of this SCC model.


\begin{theacknowledgments}
This research is supported by grants from the U.S. Department of
  Energy Office of Science, the U.S. National Science Foundation and
  the Smithsonian Institution, by NSERC in Canada, by Science
  Foundation Ireland (SFI 10/RFP/AST2748) and by STFC in the U.K. We
  acknowledge the excellent work of the technical support staff at the
  Fred Lawrence Whipple Observatory and at the collaborating
  institutions in the construction and operation of the
  instrument. 
H.~P. acknowledges support through the Young
  Investigators Program of the Helmholtz Association.
\end{theacknowledgments}


\bibliographystyle{aipproc}   

\end{document}